\documentclass[twocolumn,showpacs,prl,amsmath,amssymb]{revtex4}

\usepackage{graphicx}
\usepackage{dcolumn}
\usepackage{bm}

\begin{document}

\title{Magnetic dynamo action at low magnetic Prandtl numbers}

\author{Leonid M. Malyshkin$^{1}$} 
\email{leonmal@uchicago.edu}
\author{Stanislav Boldyrev$^{2}$}
\email{boldyrev@wisc.edu}
\affiliation{{~}$^1$Department of Astronomy \& Astrophysics,
University of Chicago, 5640 S. Ellis Ave., Chicago, IL 60637}
\affiliation{{~}$^2$Department of Physics, University of Wisconsin-Madison, 
1150 University Ave., Madison, WI 53706}


\date{\today}

\begin{abstract}
Amplification of magnetic field due to kinematic turbulent dynamo
action is studied in the regime of small magnetic Prandtl numbers.   
Such a regime is relevant for planets and stars interiors, as well as
for liquid metal laboratory experiments. A comprehensive analysis
based on the Kazantsev-Kraichnan model is reported, which establishes
the dynamo threshold and the dynamo growth rates for varying kinetic
helicity of turbulent fluctuations. It is proposed that in contrast
with the case of large magnetic Prandtl numbers, the kinematic dynamo
action at small magnetic Prandtl numbers is significantly affected by
kinetic helicity, and it can be made quite efficient with an
appropriate choice of the helicity spectrum.   

\end{abstract}


\maketitle


{\em Introduction.}---
Turbulent dynamo action is a process of amplification of a weak
magnetic field in a conducting turbulent fluid or plasma.  
It is a plausible mechanism by which cosmic 
magnetic fields were created  
\citep[e.g.,][]{lynden-bell_1994,parker_1979,moffatt_1978,kulsrud_2005,
zweibel_heiles_1997,schekochihin_2006,blackman_2006,kulsrud_2008,brandenburg_2005}. 
Dynamo action has also been studied in laboratory 
experiments \citep{spence_forest_2006,monchaux_2007,spence_etal_2009}. 

In turbulent fluids possessing velocity fluctuations in a broad range
of scales, small-scale fluctuations evolve much faster than
large-scale ones. As a result, weak magnetic fields are predominantly
amplified at the smallest scales at which the relative motion of the 
magnetic field lines is not yet significantly affected by magnetic diffusivity. 
The efficiency of the dynamo action
therefore essentially depends on fluid viscosity $\nu$ and magnetic
diffusivity $\eta$, whose ratio is characterized by the dimensionless
magnetic Prandtl number $Pm=\nu/\eta$.  

In nature, the $Pm$ number is either very large (e.g., interstellar
medium) or very small (e.g., planets, stars interiors, liquid metal
laboratory experiments). Significant interest has therefore been
devoted to the two limiting cases $Pm\gg 1$ and $Pm \ll 1$. In the
first case, the small-scale magnetic field is essentially amplified by
smooth, viscous-scale velocity fluctuations, and detailed analytical and
numerical treatment is possible, e.g., \cite{schekochihin_2002}. 
The opposite case of small $Pm$ is
understood to a much lesser extent, since in this case magnetic
fluctuations grow at the scales where the velocity field is
non-analytic, \citep[e.g.,][]{boldyrev_cattaneo_2004,eyink_2010}.  
It has been established that the dynamo action is less efficient at
$Pm \ll 1$ compared to its counterpart at $Pm\gg 1$, e.g., 
\cite{boldyrev_cattaneo_2004,ponty_2005,iskakov_2007,spence_forest_2006,reuter_2009}.   
As a result, numerical and experimental studies of low-$Pm$ dynamo
action present a significant challenge. 

In this Letter we analyze the low-Prandtl number dynamo action with the
aid of the kinematic Kazantsev-Kraichnan model
\citep{kazantsev_1968,kraichnan_1968}. The model allows one to derive
the  equations for the magnetic field correlation function, which then
can be solved numerically for extremely large, practically relevant
Reynolds numbers. We found that in contrast with the high-$Pm$ dynamo
action, low-$Pm$ dynamo action is significantly affected by {\em kinetic
helicity}. Based on our results we  propose  
that the low-$Pm$ dynamo action can be made very efficient by ensuring
that the velocity fluctuations possess large enough kinetic helicity
at the resistive scales.  \\

{\em Formulation of the model.}---The evolution of magnetic field
${\bf B}({\bf x, t})$ in magnetohydrodynamics (MHD) is described by
the induction equation 
\begin{eqnarray} 
\partial_t {\bf B}=\nabla \times ({\bf v}\times {\bf B})+\eta \nabla^2 {\bf B}, 
\label{induction}
\end{eqnarray} 
where ${\bf v}({\bf x}, t)$ is the fluid 
velocity and $\eta$ is the magnetic diffusivity. In general, the
velocity fluctuations may possess nonzero kinetic helicity, 
$H=\int {\bf v}\cdot ({\nabla \times {\bf v}})\, d^3 x\neq 0$.   

In the Kazantsev-Kraichnan model  
random turbulent velocity field is assumed to be 
homogeneous, isotropic, delta-correlated in time, and 
Gaussian with zero mean, and the covariance tensor
\begin{eqnarray} 
&\langle {v^i}({\bf x},t){v^j}({\bf x}',t') \rangle \!=\!
\kappa^{ij}({\bf x}-{\bf x}')\delta(t-t'), 
\label{V_V_TENSOR} \\
&\kappa^{ij}({\bf x})\!=\!\kappa_N 
\left(\delta^{ij}-{\hat x}^i{\hat x}^j\right)+
\kappa_L {\hat x}^i{\hat x}^j+g\epsilon^{ijk}x^{k},    
\label{KAPPA}
\end{eqnarray}
where $x=|{\bf x}|$, ${\hat {\bf x}} \equiv {\bf x}/x$,  $\langle...\rangle$ 
denotes ensemble average, 
$\epsilon^{ijk}$ is the unit anti-symmetric pseudo-tensor, and
summation over repeated indices is assumed. The first two terms in  
the right-hand side of Eq.~(\ref{KAPPA}) describe the mirror-symmetric, 
nonhelical part of the turbulence, while the last term describes 
the helical part. We assume that the velocity is incompressible, 
resulting in the relation $\kappa_N(x)=\kappa_L(x)+(x/2)(d\kappa_L/dx)$. 
The velocity field is fully specified 
by the two functions $\kappa_L(x)$ and $g(x)$, related to the kinetic
energy and  
kinetic helicity, correspondingly. The Fourier analog of
(\ref{V_V_TENSOR}, \ref{KAPPA}) is straightforward:  
\begin{eqnarray} 
& \langle {\hat v}^{i*}({\bf k},t){\hat v}^j({\bf k},t')\rangle 
= {\hat \kappa}^{ij}({\bf k})\delta(t-t'),  
\label{V_V_TENSOR_FOURIER}\\
& {\hat \kappa}^{ij}({\bf k})
=F(k)\left(\delta^{ij}-{\hat k}^i{\hat k}^j\right)+iG(k)\epsilon^{ijl}k^l, 
\label{KAPPA_FOURIER}
\end{eqnarray}
where ${\hat {\bf k}}={\bf k}/k$. The 
functions $F(k)$ and $G(k)$ are related to $\kappa_L(x)$ and $g(x)$  
by means of the three-dimensional Fourier 
transform~\citep{monin_1971}. There is a limit 
to the maximal kinetic helicity due to the Schwarz
inequality~\citep{moffatt_1978},  
\begin{eqnarray}
|G(k)|\le F(k)/k.
\label{REALIZABILITY}
\end{eqnarray}

The correlation function of a homogeneous and isotropic magnetic field can be 
written similarly to Eq.~(\ref{KAPPA}),
\begin{eqnarray}
& \langle B^i({\bf x}, t)B^j(0,t)\rangle = 
M_N\left(\delta^{ij}-{\hat x}^i{\hat x}^j\right) 
+ M_L{\hat x}^i{\hat x}^j \nonumber \\
& +K\epsilon^{ijk}x^k.
\label{B_B_TENSOR}
\end{eqnarray} 
Because the magnetic field is divergence-free, we have 
$M_N(x,t)=M_L(x,t)+(x/2)(\partial M_L/\partial x)$. The magnetic field
correlator is fully specified by the two functions,  
$M_L(x, t)$ and $K(x, t)$, which correspond to magnetic energy and 
magnetic helicity respectively. 
In the kinematic dynamo theory the Lorentz force acting on the
magnetized fluid is neglected, and the velocity  
field~(\ref{V_V_TENSOR}) is considered to be prescribed. 

It turns our that the dynamo problem can be reduced to a quantum 
mechanical ``spinor'' form with imaginary time:
\begin{eqnarray}
-\partial_t \psi^{\alpha}(x, t) ={\hat{\cal H}}^{\alpha \beta}\psi^{\beta},
\label{spinor}
\end{eqnarray}
where $\alpha=\{1,2\}$ and summation over repeated indices is assumed. 
The Hamiltonian ${\hat{\cal H}}^{\alpha\beta}$ is self-adjoint and 
it depends on kinetic energy $\kappa_L(x)$,  
kinetic helicity~$g(x)$, and magnetic diffusivity~$\eta$ \citep{boldyrev_2005}. 
The magnetic correlator functions $M_L(x, t)$ and $K(x, t)$ can then be 
expressed in terms of the two components of the function 
$\psi^{\alpha}(x, t)$. 
In the case of zero kinetic helicity, $g(x)\equiv 0$, 
Eq.~(\ref{spinor}) reduces to the Kazantsev differential 
equation~\citep{kazantsev_1968}. 

Negative eigenvalues of ${\hat{\cal H}}^{\alpha \beta}$,  
$-\lambda<0$,  correspond to exponentially 
growing magnetic fluctuations,  
$\psi^{\alpha}(x, t)\propto e^{\lambda t}$, $M_L(x, t)\propto e^{\lambda t}$ 
and $K(x, t)\propto e^{\lambda t}$.
Similar to quantum mechanics, in a general case of 
non-zero kinetic helicity, $g(0)\ne 0$, there are two types of 
eigenfunctions of Eq.~(\ref{spinor}) \citep{boldyrev_2005,malyshkin_2007}. 
First, there are unbound (spatially non-localized)
eigenfunctions which correspond to continuous spectrum
$0<\lambda\le\lambda_0$. Here  
$\lambda_0=g^2(0)/[\kappa_L(0)+2\eta]$ is the largest growth rate of
an unbound eigenmode. At large  $x$ the unbound modes  
become mixtures of cosine and sine standing waves. Second, there are
bound (spatially localized) eigenfunctions that have  
discrete spectrum $\lambda_n>\lambda_0$, $n=1,2,...$. These
eigenfunctions decay exponentially fast at large~$x$, 
\begin{eqnarray}
M_L(x, t)&\propto& x^{-2}e^{-k_rx}\cos[k_ix+\phi]\nonumber \\
K(x, t)&\propto& x^{-2}e^{-k_rx}\sin[k_ix+\psi], 
\label{large_x_asymptotic}
\end{eqnarray}
where
$k_r=\sqrt{\lambda_n-\lambda_0}/\sqrt{\kappa_L(0)+2\eta}$ and  
$k_i=\sqrt{\lambda_0}/\sqrt{\kappa_L(0)+2\eta}$  \citep{boldyrev_2005,malyshkin_2007}. 
Note that the existence of unbound and bound eigenmodes in the helical 
dynamo problem is similar to existence of ``free'' and ``trapped'' quantum 
particles traveling in a potential well.

As long as the kinetic helicity is non-zero, the unbound eigenmodes always
exist in an infinite system. However, real (astro)physical or laboratory 
systems are limited in size and they may not allow for spatially 
unbound eigenfunctions. In particular, the unbound solutions should be not 
relevant if the system size $L$ is smaller than $1/k_i\sim \kappa_L(0)/g(0)$. 
We restrict our consideration to the bound eigenmodes.  

We will assume the Kolmogorov scaling of velocity fluctuations,
$\langle|{\hat v}({\bf k},t)|^2\rangle\sim v_0^2l_0^{-2/3}k^{-11/3}$,
with the corresponding eddy turnover time  
$\tau(k)\sim v_0^{-1}l_0^{1/3}k^{-2/3}$~\citep{frisch_1995}. Here $v_0$ and 
$l_0$ are the velocity and the scale associated with the largest
eddies. The final equation (\ref{spinor}) in the Kazantsev-Kraichnan
model involves     
only the integral of the velocity correlation function over time, that
is, the turbulent diffusivity,  which in Kolmogorov turbulence scales
as $\langle|{\hat v}({\bf k},t)|^2\rangle \tau(k)\sim
v_0l_0^{-1/3}k^{-13/3}$. By requiring that the Kazantsev-Kraichnan
model have the same scaling of turbulent diffusivity, we obtain      
$\int\langle{\hat v}^{i*}({\bf k},t){\hat v}^i({\bf k},t')\rangle dt =
{\hat\kappa}^{ii}({\bf k}) \sim v_0l_0^{-1/3}k^{-13/3}$, and, therefore, 
$F(k)={\hat\kappa}^{ii}({\bf k})/2\sim v_0l_0^{-1/3}k^{-13/3}$. 
Without loss of generality we assume $l_0\sim 1$, $v_0\sim 1$, and we also choose  
\begin{eqnarray}
F(k)=Ck^{-13/3},\quad G(k)=-hF(k)/k 
\label{POWER_V}
\end{eqnarray}
for $2\le k\le k_{\rm max}$, and $F(k)=G(k)= 0$ otherwise. 
Here $C=(1/2\pi)/(2^{-4/3}-k_{\rm max}^{-4/3})\sim 1$ 
is a normalization coefficient chosen such that $\kappa_L(0)=1$; note 
that $\kappa_L(0)\sim l_0v_0$ is the turbulent diffusivity.
The maximal wavenumber is 
$k_{\rm max}\approx 2[\kappa_L(0)/\nu]^{3/4}=2\nu^{-3/4}$, 
where $\nu$ is the kinematic viscosity. For convenience, we 
specify $k_{\rm max}$, and then define the effective viscosity as
$\nu=\kappa_L(0)(2/k_{\rm max})^{4/3}=(2/k_{\rm max})^{4/3}$.
The simple choice of $G(k)$ in Eq.~(\ref{POWER_V}), with $-1\le h\le 1$ due 
to the realizability condition~(\ref{REALIZABILITY}), assures that kinetic 
helicity is present on all scales (we study the effect of helicity on 
different scales below). The helicity is maximal when $|h|=1$.
We define the Reynolds number as $Re=\kappa_L(0)/\nu=1/\nu$, 
the magnetic Reynolds number as $Rm=\kappa_L(0)/\eta=1/\eta$, and the
magnetic Prandtl number as $Pm=Rm/Re=\nu/\eta$. 

For exponentially growing  magnetic field, 
$\psi^{\alpha}(x, t)\propto e^{\lambda t}$, 
Eq. (\ref{spinor}) turns into a system of two 
ordinary differential equations with variable coefficients. 
We integrate this system numerically by the fourth-order Runge-Kutta method
on a nonuniform numerical grid, as described in~\citep{malyshkin_2007}. 
As a result, we find the growth rates and the eigenfunctions of all the bound 
and unbound modes. The bound eigenfunctions decline exponentially at infinity,  
therefore, they are set to zero at the right boundary of our 
computational interval. \\

{\em Dynamo growth rate.}---As follows from the phenomenological
discussion in the introduction, weak magnetic field is most
efficiently amplified by turbulent eddies at the resistive scale. 
Using the Kolmogorov scaling $v_l\sim
l^{1/3}$, we find from Eq.~(\ref{induction}) that the resistive scale
is $l_\eta\sim \eta^{3/4}$, and the eddy turnover time at this scale
is $\tau_\eta\sim l_\eta/v_\eta\sim \eta^{1/2}$. The dynamo growth
rate should therefore scale as $\lambda\sim 1/\tau_\eta \sim
Rm^{1/2}$, see, e.g.,
\cite{boldyrev_cattaneo_2004,iskakov_2007,reuter_2009}.   

\begin{figure}
\includegraphics[width=0.48\textwidth]{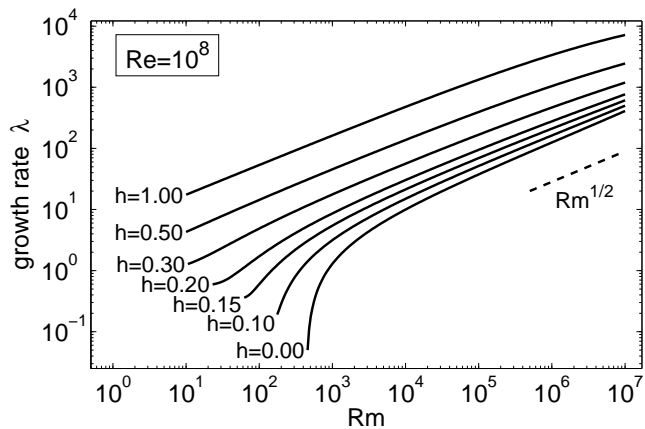}
\caption{Scaling of the dynamo growth rate vs the magnetic Reynolds
  number~$Rm$, obtained for various levels of kinetic helicity $h$. }
\label{Rm_scaling}
\end{figure}

This result is hard to check in low-$Pm$ direct numerical simulations,
since the range of  
$Rm$ numbers available in such simulations is significantly limited
(see, however, \cite{haugen_2004}, where some indication of such
scaling was found in the case $Pm\sim 1$). Here we check this scaling
in the Kazantsev-Kraichnan model. We solve the model for $Re\sim
10^8$, and for the magnetic Reynolds numbers up to $Rm\sim 10^7$. The
results are presented in Fig.~(\ref{Rm_scaling}).  

The phenomenological scaling $\lambda\sim Rm^{1/2}$ is clearly seen in
this plot for the zero helicity case $h=0$
\footnote{An earlier attempt to derive such a scaling from
  the Kazantsev-Kraichnan model produced a different result $\lambda
  \sim \log(Rm)$ \cite{rogachevskii_1997}. The discrepancy is 
  a consequence of an incorrect asymptotic matching procedure employed
  in~\cite{rogachevskii_1997} for establishing the dynamo 
  threshold and calculating the dynamo growth rate.
  }.   
Moreover, we find that the scaling is very close to $\lambda\sim Rm^{1/2}$ 
even when kinetic helicity is non-zero.  We also observe that for moderate helicities the asymptotic scaling
$\lambda\sim Rm^{1/2}$ is established only at very large $Rm$
numbers. We therefore expect that observation of such a scaling will
present significant challenge for numerical simulations and for
laboratory liquid-metal experiments.  On a more optimistic note, we
observe that kinetic helicity facilitates dynamo action; this effect
will be more evident in our discussion of the dynamo threshold in the
next section.\\

\begin{figure}
\includegraphics[width=0.48\textwidth]{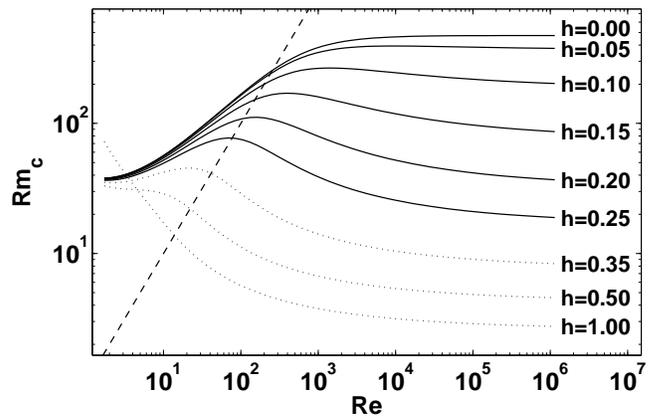}
\caption{Stability curves   
for different values of kinetic helicity~$h$ for the box size $L=1/0.3$. 
The curves correspond to the bound modes, that is, the modes with
$k_r>1/L$. For  $h>\kappa(0)/g(0)\sim 0.3 $, the unbound modes can
also be excited in the box. In these cases,
the presented $Rm_c$ for the bound modes (shown by dotted lines) give 
the upper boundary for the true $Rm_c$.  The dashed line 
corresponds to $Pm=1$.}
\label{FIGURE_1}
\end{figure}

{\em Dynamo threshold.}---A crucial question of whether the dynamo
action can be observed in a particular system is related to the
question of dynamo threshold. We address this question by solving the
model for the system size $L=1/0.3$ in units of the largest turbulent  
eddy size $l_0\sim1$. We consider the Reynolds numbers ranging from
$Re\approx 1$ to $Re\approx 10^6$, corresponding to the 
maximal wavenumber ranging
from $k_{\rm max}=3$ to $k_{\rm max}=7\times 10^4$. For each $Re$ (and
the corresponding $k_{\rm max}=2Re^{3/4}$) we find the critical 
magnetic Reynolds number $Rm_c$ for 
which there is at least one growing bound mode with $k_r>1/L$. 
The resulting $Rm_c$ values are shown for 
different values of kinetic helicity $h$ in 
Figure~\ref{FIGURE_1}. The dashed line corresponds to 
$Rm=Re$, that is, $Pm=1$. 
We now make the following two important observations. 

First, in the case when kinetic helicity is zero, $h=0$, the 
critical magnetic Reynolds numbers $Rm_c$ are considerably higher in 
low-$Pm$ turbulence, 
than in high-$Pm$ turbulence. 
This result is in agreement with previous analytical and numerical studies 
of low-$Pm$ dynamo in various geometries
\citep{boldyrev_cattaneo_2004,
  ponty_2005,iskakov_2007,reuter_2009}. The situation changes
drastically when $h\neq 0$.  
We find that the critical $Rm$ 
in low-$Pm$ turbulence is very sensitive to the amount of kinetic
helicity, while it is practically   
independent of kinetic helicity in high-$Pm$ turbulence. We will
analyze this  important property in more detail below. 

The second observation is that the stability curve $Rm_c(Re)$ ceases
to be monotone as kinetic helicity increases. The curve peaks around
$Rm\sim Re$, before it declines and eventually flattens at $Re\gg
Rm$. This overshooting effect seems to be a robust feature of  recent
numerical simulations in periodic boxes and in spherical geometry
\cite{ponty_2005,iskakov_2007,reuter_2009}, although its origin has
not been understood. Here we propose that the overshooting is a
consequence of kinetic helicity of the
velocity fluctuations. Even if kinetic helicity is zero overall (as,
e.g., is \cite{iskakov_2007}), it can  possess spatial variations,
which can affect small scale dynamo action.\\   
\begin{figure}
\includegraphics[width=0.48\textwidth]{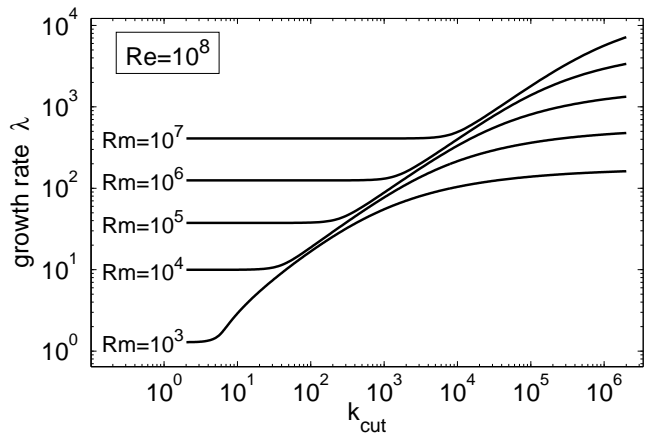}
\caption{Dynamo growth rates for varying spectrum of magnetic
  helicity: $G(k)=0$ at $k> k_{cut}$ and  $G(k)=-F(k)/k$ at $k \leq k_{cut}$.}
\label{FIGURE_3}
\end{figure}

{\em Role of kinetic helicity.}---In order to elucidate the role of
kinetic helicity in low-$Pm$ dynamo, we study at what scales the
influence of kinetic helicity is most significant. We set $Re=10^8$,
and consider several magnetic Reynolds numbers well above the
dynamo threshold, where the growth rates are given by
Fig.~(\ref{Rm_scaling}) in the fully helical case ($h=1$). However, 
instead of considering the fully helical case, we now modify the  
helicity spectrum  $G(k)$ in Eq.~(\ref{POWER_V}) as follows. We set
$G(k)=0$ at large wavenumbers $k> k_{cut}$ and 
keep $G(k)=-F(k)/k$ at $k \leq k_{cut}$.  We then study how the dynamo
growth rate changes as we change the cutoff $k_{cut}$. The result is
presented in Fig.~(\ref{FIGURE_3}).  

We observe that the presence of large-scale kinetic helicity (small
$k_{cut}$) does not affect the  
growth rate. Varying kinetic helicity at very small scales (large
$k_{cut}$) does not affect the growth rate either. Since the only
characteristic scale present in the kinetic inertial interval is the
resistive scale, $k_\eta\sim Rm^{3/4}$, we suggest that kinetic
helicity significantly affects the dynamo action only when it is
present at the resistive scales of magnetic fluctuations. This
statement is physically reasonable since, as was explained earlier, it
is the resistive scales where the growing magnetic field is
concentrated in the kinematic dynamo action. The scaling of the
critical cut-off number $k_{cut} \sim Rm^{3/4}$ is indeed consistent
with the behavior observed in Fig.~(\ref{FIGURE_3}). \\ 

{\em Conclusions.}---We studied kinematic dynamo action in low-$Pm$
turbulence in the framework of the solvable  Kazantsev-Kraichnan model
\cite{kazantsev_1968,kraichnan_1968}. Based on the obtained results we
proposed the explanations for somewhat puzzling results of recent direct
numerical simulations of low-$Pm$ dynamo action, concerning the dynamo
threshold, and the scaling of the dynamo growth rate with the magnetic
Reynolds number. We also proposed   
that the low-$Pm$ turbulent dynamo action can be made very efficient
by ensuring that the velocity fluctuations possess large enough
kinetic helicity at the resistive scales.

\acknowledgments
We thank Fausto Cattaneo and Joanne Mason for helpful discussions. 
This work was supported by the NSF Center for Magnetic 
Self-Organization at the University of Chicago under the NSF award PHY-0821899 (LMM), and  
by the US DoE grant DE-FG02-07ER54932  
and the US DoE Early Career Award DE-SC0003888 (SB).




\begin{thebibliography}{}

\bibitem[Kulsrud(2005)]{kulsrud_2005}
R.~M.~Kulsrud, R.~M., {\em Plasma Physics for Astrophysics} 
(Princeton Univ.~Press, Princeton, NJ, 2005).

\bibitem[Brandenburg \& Subramanian(2005)]{brandenburg_2005} 
A.~Brandenburg and K.~Subramanian, Phys. Rep. {\bf 417}, 1 (2005).

\bibitem[Blackman \& Ji(2006)]{blackman_2006} 
E.~Blackman and H.~Ji, Mon. N. R. Astron. Soc. {\bf 369}, 1837 (2006).

\bibitem[Kulsrud \& Zweibel(2008)]{kulsrud_2008}
R.~M.~Kulsrud and E.~G.~Zweibel, 
Rep. Prog. Phys. {\bf 71}, 046901 (2008).

\bibitem[Lynden-Bell(1994)]{lynden-bell_1994}
D.~Lynden-Bell (ed.), {\em Cosmical Magnetism} 
(Kluwer, Dordrecht, Holland, 1994).

\bibitem[Moffatt(1978)]{moffatt_1978}
H.~K.~Moffatt,
{\em Magnetic Field Generation in Electrically Conducting Fluids} 
(Cambridge University Press, Cambridge, England, 1978).

\bibitem[Parker(1979)]{parker_1979}
E.~N.~Parker, {\em Cosmical Magnetic Fields} 
(Clarendon Press, Oxford, England, 1979).

\bibitem[Schekochihin \& Cowley(2006)]{schekochihin_2006}
A.~A.~Schekochihin and S.~C.~Cowley, 
Astron. Nachr. {\bf 327}, 599 (2006).

\bibitem[Zweibel \& Heiles(1997)]{zweibel_heiles_1997}
E.~G.~Zweibel and C.~Heiles, 
Nature {\bf 385}, 131 (1997).

\bibitem[Monchaux, {\it etal.}(2007)]{monchaux_2007}
R.~Monchaux et al., 
Phys. Rev. Lett. {\bf 98}, 044502 (2007).

\bibitem[Spence {\it et al.}(2006)]{spence_forest_2006}
E.~J.~Spence {\it et al.}, 
Phys. Rev. Lett. {\bf 96}, 055002 (2006).

\bibitem[Spence {\it et al.}(2009)]{spence_etal_2009} 
E.~J.~Spence, K.~Reuter and C.~B.~Forest, 
Astrophys. J. {\bf 700}, 470 (2009).

\bibitem[Schekochihin, Boldyrev \& Kulsrud(2002)]{schekochihin_2002} 
A.~A.~Schekochihin, S.~Boldyrev and R.~Kulsrud, 
Astrophys. J. {\bf 567}, 828 (2002). 

\bibitem[Boldyrev \& Cattaneo(2004)]{boldyrev_cattaneo_2004}
S.~Boldyrev and F.~Cattaneo, 
Phys. Rev. Lett. {\bf 92}, 144501 (2004).

\bibitem[Eyink \& Neto(2010)]{eyink_2010} 
G.~L.~Eyink and A.~F.~Neto, 
New J. Phys. {\bf 12}, 023021 (2010).  

\bibitem[Iskakov {\it et al.}(2007)]{iskakov_2007}
A.~B.~Iskakov {\it et al.}, 
Phys. Rev. Lett. {\bf 98}, 208501 (2007).

\bibitem[Reuter(2010)]{reuter_2009} 
K.~Reuter, ``Numerical investigation of turbulent dynamo 
excitation in a spherical MHD system", Ph.~D. Thesis 
(Max-Planck-Institut f\"ur Plasmaphysik, Garching, 2010).

\bibitem[Ponty {\it et al.}(2005)]{ponty_2005}
Y.~Ponty {\it et al.}, 
Phys. Rev. Lett. {\bf 94}, 164502 (2005).

\bibitem[Kazantsev(1968)]{kazantsev_1968}
A.~P.~Kazantsev, 
JETP {\bf 26}, 1031 (1968).

\bibitem[Kraichnan(1968)]{kraichnan_1968}
R.~H.~Kraichnan, 
Phys. Fluids {\bf 11}, 945 (1968).

\bibitem[Monin \& Yaglom(1971)]{monin_1971}
A.~S.~Monin and A.~M.~Yaglom, 
{\em Statistical Fluid Mechanics} 
(MIT Press, Cambridge, MA, 1971).

\bibitem[Boldyrev, Cattaneo \& Rosner(2005)]{boldyrev_2005}
S.~Boldyrev, F.~Cattaneo and R.~Rosner, 
Phys.~Rev.~Lett. {\bf 95}, 255001 (2005).

\bibitem[Malyshkin \& Boldyrev(2007)]{malyshkin_2007}
L.~Malyshkin and S.~Boldyrev, 
Astrophys. J. Lett. {\bf 671}, L185 (2007).

\bibitem[Frisch(1995)]{frisch_1995}
U.~Frisch, {\em Turbulence} 
(Cambridge Univ.~Press., Cambridge, England, 1995).

\bibitem[Haugen, Brandenburg \& Dobler(2004)]{haugen_2004}
N.~E.~L.~Haugen, A.~Brandenburg and W.~Dobler,  
Phys. Rev. E {\bf 70}, 016308 (2004).

\bibitem[Rogachavskii \& Kleeorin(1997)]{rogachevskii_1997} 
I.~Rogachevskii and N.~Kleeorin, 
Phys. Rev. E {\bf 56}, 417 (1997).













\end{thebibliography}
\end{document}